\documentclass[graybox]{svmult}

\usepackage{cite}
\usepackage{textcomp}
\usepackage{xcolor}
\def\BibTeX{{\rm B\kern-.05em{\sc i\kern-.025em b}\kern-.08em
    T\kern-.1667em\lower.7ex\hbox{E}\kern-.125emX}}
\usepackage{booktabs} % For formal tables
\usepackage{subfig}
\usepackage{mathrsfs}
\usepackage{amsmath,amssymb,amsfonts}

\usepackage{algorithm}
\usepackage[noend]{algpseudocode}

\usepackage{mathptmx}       % selects Times Roman as basic font
\usepackage{helvet}         % selects Helvetica as sans-serif font
\usepackage{courier}        % selects Courier as typewriter font
\usepackage{type1cm}        % activate if the above 3 fonts are
\usepackage{makeidx}         % allows index generation
\usepackage{graphicx}        % standard LaTeX graphics tool
\usepackage{multicol}        % used for the two-column index
\usepackage[bottom]{footmisc}% places footnotes at page bottom

\graphicspath{{./Images/}}

\makeindex             % used for the subject index
\begin{document}

\title*{\textsf{Multi-Net}: A Scalable Multiplex Network Embedding Framework}
% Use \titlerunning{Short Title} for an abbreviated version of
% your contribution title if the original one is too long
\author{Arunkumar Bagavathi and Siddharth Krishnan}
% Use \authorrunning{Short Title} for an abbreviated version of
% your contribution title if the original one is too long
\institute{Arunkumar Bagavathi \at University of North Carolina, Charlotte, North Carolina, USA \email{abagavat@uncc.edu}
\and Siddharth Krishnan \at University of North Carolina, Charlotte, North Carolina, USA \email{skrishnan@uncc.edu}}
%
% Use the package "url.sty" to avoid
% problems with special characters
% used in your e-mail or web address
%
\maketitle

\abstract{
Representation learning of networks has witnessed significant progress in recent times. Such representations have been effectively used for classic network-based machine learning tasks like node classification, link prediction, and network alignment. However, very few methods focus on capturing representations for \emph{multiplex or multilayer} networks, which are more accurate and detailed representations of complex networks. In this work, we propose \textsf{Multi-Net} a fast and scalable embedding technique for multiplex networks. \textsf{Multi-Net}, effectively maps nodes to a lower-dimensional space while preserving its neighborhood properties across all the layers. We utilize four random walk strategies in our unified network embedding model, thus making our approach more robust than existing state-of-the-art models. We demonstrate superior performance of \textsf{Multi-Net} on four real-world datasets from different domains. In particular, we highlight the uniqueness of \textsf{Multi-Net} by leveraging it for the complex task of network reconstruction.
}

\section{Introduction}
%\vspace{-0.2in}
	Networks are ubiquitous and are popular mathematical abstractions to analyze data, particularly web data. The natural interconnectedness that web data, like social and information networks, presents make networks a natural metaphor to understand the relationships exhibited in the high-volume multi-modal nature of data. Thus most analytics research for networks such as \textit{link prediction}~\cite{backstrom2011} or \textit{node classification}~\cite{bhagat2011} involving machine learning, require features that are descriptive and discriminative. Thus it is important to translate complex graph data into a set of attributes through creative feature engineering, which can often be difficult to compute, especially for high volume of data. 

	The significant research thrust and recent successes in word and document embeddings (word2vec and doc2vec)~\cite{mikolov2013,le2014} have motivated researchers to explore analogous representations in related fields, for example networks. Recent works like \textsf{node2vec, DeepWalk} (see related works) have explored dynamic feature representations of a large scale network to a low-dimensional vector space. These low-dimensional features have been effectively used in network specific analytics like \textit{link prediction}~\cite{backstrom2011} or \textit{node classification}~\cite{bhagat2011}.
\begin{figure}
\centering
\includegraphics[scale=0.2]{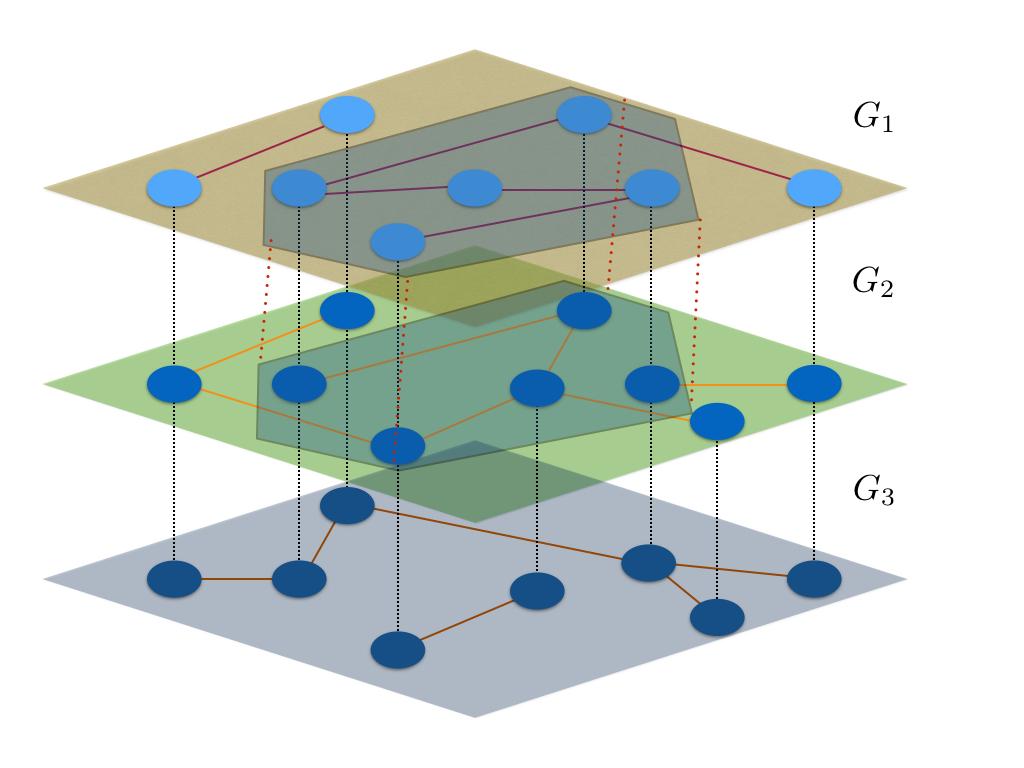}
\caption{Example of a multiplex network \textit{MG} with multiple networks \textit{$G_1,G_2,G_3$}. Each layer can be either different media like Twitter, Facebook, and LinkedIn or different interaction types like retweet, mention and reply.}
\label{fig:multilayer}
\end{figure}
	However, large scale networks such as social networks and human brain tissue networks, in the real world, exist with multiple layers of embedded or independent networks. For example, in case of social networks, a user is usually present in multiple social media platforms like Twitter, Facebook, Youtube, etc. and exhibits different level of interaction and behavior in each network as given in Figure~\ref{fig:multilayer}. Although, few existing models focus on representing nodes based on multiplex networks~\cite{zitnik2017,zhang2018}, these representations can be made more robust by exploring random walk techniques that are dedicated for multiplex networks. Using these techniques, an embedding technique that accounts for the multi-layered existence and interaction of the nodes will equip learning algorithms with a richer vocabulary for learning tasks like link prediction across different networks. With this multiplex node embeddings, we explore a network reconstruction task. For example, by learning a node's interaction patterns in the \emph{retweet} and \emph{mentions} network, can we predict the underlying friend-follower network structure? This question is one of the motivating examples for our study and presents an interesting use-case of \textsf{Multi-Net}.\\

\subsection*{Summary of contributions}
Our major contributions in this paper include:
\begin{itemize}
	\item \textbf{How a multiplex network can be efficiently explored to create a node context?} We provide multiple random walk strategies that effectively navigate through the layered structure of the networks. The crux of such strategies lies in devising a random walker's transition probability to account for in-layer traversal and a switching probability for cross-layered movement.
	\item \textbf{How large scale multiplex networks can be represented in a lower dimensional space?} We propose \textsf{Multi-Net} a fast, scalable, deep network embedding algorithm that preserves the cross-layered neighborhood of a node across its various interaction patterns.
	
	%\item \old{\textbf{How to construct a friends/follower network from the textual data?} By mining the mentions, retweet, and reply networks which can be easily obtained from tweet content and its associated metadata, we use \textsf{Multi-Net} to effectively learn distributed feature representations of the nodes involved and infer the friend-follower structure of the Twitter network. We outperform state-of-the-art methods by approximately 7\%. (See Sec. Experimental Results)}
	\item \textbf{How to reconstruct a layer in a multiplex network using Multi-Net framework?} By extracting distributed feature representations of a node in a multilayered neighborhood preserving context and its structural arrangement in $\mathscr{L}-1$ layers, we use \textsf{Multi-Net} to effectively infer the complete structure of the $\mathscr{L}^{th}$ network layer. Our experiments show that our models outperform recently proposed multiplex network embedding methods \textsf{OhmNet}~\cite{zitnik2017} and \textsf{MNE}~\cite{zhang2018} by approximately 11\% accuracy. To our knowledge, this is the first work to evaluate multiple random walks for multiplex network embedding and compare against existing models.

	%\item We propose a random walk procedure for multilayer network 
	%\item We show how these random walks can help in learning features of multilayer networks without maintaining any hierarchy for social networks
	%\item We propose a scalable algorithm for feature learning on multilayer network that optimizes neighborhood preserving objective function
	%\item We provide insights on how the representations learned from multilayer networks can be useful in multitude of real world problems
	%\item We evaluate learned multilayer feature representations, from Twitter's retweet, reply and mentions network, over a friend or follower prediction task
\end{itemize}

%\vspace{-0.25in}
\section{Related work}
%\vspace{-0.2in}
Feature learning is one of the sustained and prominent topics in the machine learning group. Feature representations are most required to study the properties and do more accurate predictions or forecasting on large scale social/other networks. Over a decade human engineered feature representations, based on network's structural and temporal properties, has been widely used in multiple tasks like node classification \cite{henderson2011}, characterizing information cascades \cite{krishnan2016}. Although these works give promising results, they are task specific. The Skip-gram model proposed by Mikolov et al.~\cite{mikolov2013} has spurred lot of activity among research communities to automatically learn features from data, like networks. The end goal of all network embedding methods is to learn an encoding for the network nodes that effectively captures crucial properties of a node such as their neighborhood connectivity, degree, and position. Since such methods group together nodes that share same properties, these techniques are being used for variety of tasks like node and link prediction~\cite{liu2010,perozzi2014,grover2016} , node clustering \cite{leskovec2012} and community detection \cite{pons2006}.

Network embedding literature can take two-fold. The first method is using matrix factorization techniques. With this method, the entire graph is represented as an adjacency matrix where each cell in the matrix represent node relationship. This matrix is processed and projected into a lower dimensional space \cite{kramer2014}. This method comes with a cost of lower scalability and it does not capture structural properties of a node in the network. Neural network based network embedding can overcome the above mentioned limitations of matrix factorization approach. Nodes in the network can be organized in such a way that the neural network based Skip-gram model can be applied to learn features by optimizing neighborhood preserving objective. The most efficient method to gather network neighborhood is to do random walks \cite{backstrom2011}. The two most popular methods that uses random walks and Skip-gram model to learn features of large-scale networks are \textsf{DeepWalk} \cite{perozzi2014} and \textsf{Node2Vec} \cite{grover2016}.

\begin{table}
\centering
\caption{Summary of notations}
\label{tab:paper_notations}
\begin{tabular}{|p{3cm}|p{5cm}|}
\hline
Notation & Description \\
\hline
$MG$ & Multiplex network \\
\hline
$\mathscr{L}$ & number of layers in $MG$ \\
\hline
$\mathscr{V}$ & non-empty set of vertexes in $MG$ \\
\hline
$\mathscr{E}$ & non-empty set of edges in $MG$ \\
\hline
\textit{d} & size of feature dimension of a node \\
\hline
$C_u$ & Set of neighborhood nodes of node $u$ \\
\hline
$w_{ij}^{\alpha\beta}$ & transition probability to move from node \textit{i} in layer $\alpha$ to \textit{j} in layer $\beta$ \\
\hline
$\mathbb{A}$ & adjacency matrix \\
\hline
$D^{\alpha\alpha}$ & distance matrix(shortest distance to travel between nodes) of a network $\alpha$ \\
\hline
$s_i$ & strength of node $i$ defined as $\sum_{j,\beta} \big(a_{ij}^{\alpha\alpha}+D_{ii}^{\alpha\beta}\big)$ with accordance with $\mathscr{L}$ layers \\
\hline
$s_{max}$ & $max_{i,\alpha}s_i^\alpha$  \\
\hline
$\mathbb{S}_{i,\alpha}$ & $\sum_{j}^{} a_{ij}^{\alpha\alpha}$  \\
\hline
\hline

\end{tabular}
\end{table}

Although predictions and forecastings based on network embeddings from the above methods gives promising results, the real world networks are complex and available in multiple forms. Some of such complex networks include temporal networks \cite{rossi2013}, sign network \cite{leskovec2010} and multi-layer network \cite{boccaletti2014}. Learning representations from such networks can improve prediction and forecasting power of a learning algorithm. We focus on feature representations on a special kind of multilayer network called \textit{multiplex network}. The key challenge involved in learning features from a multiplex network is to consider learning from multiple layers or networks to give a single representation. The very naive approach of extracting features from the multilayer network is to merge all networks into a single network~\cite{boccaletti2014,loe2015}. Combining multiple networks into a single network may disturb the topological order or properties of nodes with respect to each network in the multiplex network~\cite{de2016}. To overcome this limitation, a hierarchy based model has been proposed to learn features from multilayer networks that uses \textsf{Node2Vec} \cite{grover2016} to get embeddings of multiple networks and a hierarchy to combine such embeddings \cite{zitnik2017}. The drawback of this approach is that it relies on a hierarchy of layers of networks to combine the results. However, in the real world multilayer networks such as social networks, such hierarchy does not exist and manually designing such hierarchies may not represent the network to the highest precision. Recently a Scalable Multiplex Network Embedding(MNE)~\cite{zhang2018} has been introduced, which has performance overload since it extracts additional embeddings of nodes from multiple layers in the network. In this paper, we propose a simple \textit{random walk} procedure, inspired from \cite{guo2016}, to organize nodes across multiple networks and use such sequence of nodes to learn their low-dimensional features.
%\vspace{-0.25in}
\section{Problem Formulation}
%\vspace{-0.2in}

Our goal in this paper is to learn low-dimensional node features represented as continuous vectors, without extracting them manually, from a multiplex network. Table~\ref{tab:paper_notations} gives a summary of all notations used in this paper. More formally,

\textbf{Multiplex network:} A multiplex network is a tuple
\begin{center}$MG=\big(V_i, E_i\big)_{i=1}^\mathscr{L}$,\end{center}where $V_i$ and $E_i$ are the set of vertexes and edges at layer $i$ respectively and $\mathscr{L}$ is the number of layers. Informally, \textit{MG} is a collection of networks  ${G_1,G_2, \ldots G_\mathscr{L}}$. We make an assumption for the node set $\bigcap\limits_{i=1}^\mathscr{L} V_i$, for all layers $i \in \mathscr{L}$, to have non-null intersection, to learn representation of each node across multiple layers.\\\\

% \mlg{Why is the intersection of all $V_i$ required to be non-empty? Where is this hypothesis used? This condition does not appear in the formal definition of multilayer network}

\textbf{Problem Statement:}\\ \textit{Given:} A multiplex network \textit{MG} with a subset of nodes exist across multiple layers with different layer-level neighborhood topologies, random walk length $\textit{l}$, number of walks per node $\textit{n}$ and a positive embedding size \textit{d}; 
\\\textit{Aim:} To learn distributed feature representations of nodes of multiplex network ($\mathscr{V}$) from all layers of the network.

\begin{figure}
\centering
\includegraphics[scale=0.5]{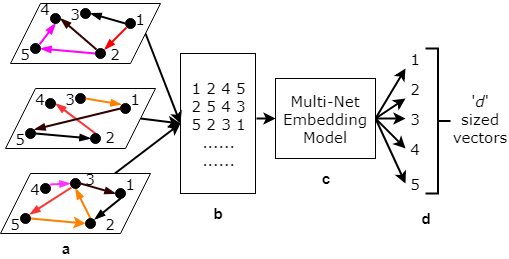}

\caption{Multi-Net methodology pipeline. (a) Proposed random walks on all nodes of across multiple network layers, (b) Sequence random walked nodes for each node, (c) Node sequences are given as inputs to the proposed embedding model, (d) Output features as \textbf{d}-dimension vectors from the embedding model}
\label{fig:multi-net-pipeline}
\end{figure}

More intuitively, consider a multiplex network $MG=\big(V_i, E_i\big)_{i=1}^\mathscr{L}$ with $\mathscr{L}$ layers of networks, where $\mathscr{V}$ is the intersection of nodes $\bigcap\limits_{i=1}^\mathscr{L} V_i$ from all $\mathscr{L}$ layers. The figurative description of our proposed approach on node embeddings from multiplex networks is given in Figure~\ref{fig:multi-net-pipeline}. For the multiplex network, we proceed with following steps:
\begin{itemize}
	%\item Given a fact that in multiplex network scenario, a set of nodes (\textit{v};\textit{v} $\subset V$) can exist in multiple layers and each node can be represented as a vector of size $|V|$. Since real world networks can have over millions of nodes ($|V| > 1M$), a machine learning algorithm cannot perform efficiently on such a high dimensional space. Thus we need a low-dimensional representation for such nodes in the multiplex network
	
	% \mlg{please clarify the role of the context window. It is not used in the learning objective, nor in the definition of the random walk, and it does not appear in Algorithm 1.}
	\item Given an embedding size \textit{d}, where $d < |\mathscr{V}|$, we first  extract the neighborhood of each node using \textit{random walks}. For each layer in the multiplex network, the random walk collects neighborhood of size $\textit{l}$ for every node. Due to the approximate nature of random walks, we run $\textit{n}$ random walks on each node to collect its neighborhood.
	%within a context window($w$)
	\item Having obtained node contexts, we learn a \textit{d-dimensional} feature vector of each node$\big($where $d<|\mathscr{V}|\big)$ using an optimization that maximizes the likelihood of neighbors of a node across $\mathscr{L}$ layers
\end{itemize}  
%\vspace{-0.2in}
\section{Methodology}

%\begin{center}
%\end{center}
%\vspace{-0.2in}

We construct our problem as a maximum likelihood optimization problem, similar to earlier works like \textsf{word2vec}~\cite{mikolov2013}, \textsf{node2vec}~\cite{grover2016} and \textsf{OhmNet}~\cite{zitnik2017}. We frame our method in such a way that it can be applied to (un)directed and (un)weighted networks also. Our method is a simple first step in a series of work to perform feature representation in a multiplex setup without requiring any hierarchy, like in~\cite{zitnik2017}, but the node representations still capture their neighborhood properties. Since a node can exist in many layers of a multiplex network, we assume that a node $v$ in layer $i$ can be related to neighbors of $v$ from other layers such as $j,k,l,\dots$ based on a probability ratio $p$.

%A multiplex network is a general term that represents several networks or layers as a single network without merging them together as a single network.
% Thus, a multilayer network can take a general form of $MG=\big(V_i, E_i\big)_{i=1}^\mathscr{L}$, where $\mathscr{L}$ is the number of layers and $V_i$ and $E_i$ is a set of nodes and edges respectively of a network $MG$.

%Our method is opposed to hierarchy based method like \cite{zitnik2017} and other multiplex network embedding method~\cite{zhang2018}, since such a hierarchy exists only in a limited scenario like biology and chemistry. Our method is a simple first step in a series of work to perform feature representation in a multiplex setup without requiring any hierarchy but still the node representations covers their neighborhood properties. Since a node can exist in multiple layers of networks, we assume that a node $v$ in layer $i$ can be related to neighbors of $v$ from other layers such as $j$ based on a probability ratio $p$.

%\vspace{-0.25in}
\subsection{Learning objective}
%\vspace{-0.2in}
In simple terms, we define our feature learning as a mapping function $f: \mathscr{V} \rightarrow \mathbb{R}^d$ that map all nodes($\mathscr{V}$) from all layers of networks $MG_1,MG_2, \ldots MG_\mathscr{L}$ to a \textit{d-dimensional} feature space, where $d < |\mathscr{V}|$, which can be used for prediction or forecasting tasks. For each node($u$) in the given multiplex network $MG$, where $u \in \mathscr{V}$, we define its multi-layer network neighborhood($C_u$). We aim to optimize the objective function, similar to the one given in \cite{mikolov2013,grover2016}, that maximizes the following log-likelihood for neighborhood $C_u$ of a node \textit{u} across all layers in the multiplex network as given in Equation~\ref{eq:likelihood}
%\mlg{the set J in equation (1) is not defined. Is J = $U_i V_i$ ? What’s the difference from V?}

\begin{equation}
\label{eq:likelihood}
	\begin{aligned}
		& \underset{\textit{f}}{\text{\textit{max}}}
		& & \sum_{i \in \mathscr{L}}^{} \sum_{u \in i}^{} log Pr \big( \big(C_u\big) | f\big(u\big) \big)
	\end{aligned}
\end{equation}

%\mlg{$C_u$ is defined as the neighborhood of u across all layers. If this is the case then the resulting optimization problem can be solved without the aid of the random walk.} where, $J= \bigcup\limits_{i=1}^{\mathscr{L}}$ %s\bigcup_{j=1}^{V_i} v_j$ 
and $C_u$ is the neighborhood of node $u$ across all $\mathscr{L}$ layers. We define the neighborhood $C_u$ of a node $u$ sample of neighbors obtained while traversing through multiplex network random walks.
The probability $Pr \big( \big(C_u\big) | f\big(u\big) \big)$ can be defined as visiting a neighbor $x\|x \in C_u$ is independent of visiting other neighbors $C_u - x$ of the node $u$, given a feature representation of node $u$ is given in Equation~\ref{eq:prob} 

\begin{equation}
\label{eq:prob}
	\begin{aligned}
		Pr \big( \big(C_u\big) | f\big(u\big) \big) = \prod\limits_{v \in C_u} Pr \big( v | f \big( u \big) \big)
	\end{aligned}
\end{equation}

Further, we model $Pr \big( v | f \big( u \big) \big)$ as a softmax function of a dot product of node pair features as given in Equation~\ref{eq:softmax}. 
%\mlg{Throughout the paper: spacing before parentheses is inconsistent.}

\begin{equation}
\label{eq:softmax}
	\begin{aligned}
		Pr \big( v | f \big( u \big) \big) = \frac{exp\left(f\left(v\right).f\left(u\right)\right)}{\sum\limits_{v \in \mathscr{V}} exp\left(f\left(v\right).f\left(u\right)\right)}
	\end{aligned}
\end{equation}

%where $V$ is the universe of all nodes across $\mathscr{L}$ layers in the multiplex network.

%\vspace{-0.25in}
\subsection{Random walks}
\subsubsection{Simple random walks}
In general, we simulate random walks of length $l$ on a set of vertices in a network $G= \big( V,E \big)$ to get a sequence of nodes. Consider a random walker is in the $n^{th}$ node($v_n$) of a random walk, starting at a pre-defined node $v_0$. The random walker can move to the next node($v_{n+1}$) based on the following probability:

\begin{center}
$P \big( v_{n+1}  =  j \ | \ v_{n}  =  i \big) \ = \ \left\{
                \begin{array}{ll}
                  \frac{w_{ij}}{c} \ if \ e_{i,j} \in E\\
                  0 \ Otherwise
                \end{array}
              \right.$
\end{center}
              
where, $e_{i,j}$ is (un)directed edge between nodes $i$ and $j$, $w_{ij}$ is the transition probability for moving from node $i$ to $j$ and $c$ is a normalizing constant.

\subsubsection{Random walks on multiplex network}
%\subsubsection{Simple random walks}
%In general, we simulate random walks of length $l$ on a set of vertices in a network $G= \big( V,E \big)$ to get a sequence of nodes. Consider a random walker is in the $n^{th}$ node($v_n$) of a random walk, starting at a pre-defined node $v_0$. The random walker can move to the next node($v_{n+1}$) based on the following probability:

%\begin{center}
% $P \big( v_{n+1}  =  j \ | \ v_{n}  =  i \big) \ = \ \left\{
 %               \begin{array}{ll}
 %                 \frac{w_{ij}}{c} \ if \ e_{i,j} \in E\\
 %                 0 \ Otherwise
 %               \end{array}
 %             \right.$
%\end{center}
              
%where, $e_{i,j}$ is (un)directed edge between nodes $i$ and $j$, $w_{ij}$ is the transition probability for moving from node $i$ to $j$ and $c$ is a normalizing constant. 

%\subsubsection{Multilayer random walks}
%\vspace{-0.2in}
Even though varieties of random walks have been proposed \cite{grover2016} for single layer networks, only few exists for multiplex networks \cite{guo2016} \cite{sole2016}. We use three random walks procedures from the literature~\cite{guo2016,sole2016} defined for multiplex networks: \textit{Classical, Diffusive} and \textit{Physical} random walks. Furthermore, we add our multi-net random walk method to the existing multilayer random walk procedures and compare results in all experiments. Table~\ref{tab:random_walks} gives transition probabilities defined for each random walk methods. We define the following formula to sample the next neighbor either from the current layer or from another layer for the multi-net random walk in Equation~\ref{eq:mult_rand}
%\mlg{Since the next layer is chosen according to the uniform distribution, beta is just a constant multiplicative term in all transition probabilities and produces no effect (since probabilities are normalized via the normalization constant c).}
%\mlg{In the current form of the algorithm the additional assumption that all the vertices have a nonzero out-degree in each of the layers is needed.}

\begin{equation}
\label{eq:mult_rand}
\begin{aligned}
P \big( v_{n+1} = j_\beta \ | \ v_{n} = i_\alpha \big) \ = \ \left\{
                \begin{array}{ll}
                  \frac{w_{ij}^{\gamma\gamma}}{c} \ if \ e_{i_\gamma,j_\gamma} \in \mathscr{E}_\gamma \ , \ i \neq j \\
                  0 \ if \ i = j
                \end{array}
              \right.
\end{aligned}
\end{equation}

where, $\gamma \in \{\alpha,\beta\}$ and $w_{ij}^{\gamma\gamma}$ is the transition probability of moving from node $i$ to $j$ either within the same layer($\alpha$) or to a different layer($\beta$) as given in Table~\ref{tab:random_walks}. Here the layers $\big\{\alpha,\beta,\gamma\big\} \subset \big\{1,2, \ldots \mathscr{L}\big\}$. Based on the above transition probability function, a random walker can walk from one node to another either within the network or another layer without maintaining any hierarchy of layers. We calculate transition probabilities based on node degree and the transition between nodes happen in uniform distribution of \textit{$\mathscr{L}$} layers.

\begin{center}
\begin{table}
\centering
\caption{Transition probabilities of nodes in for random walks in the multiplex network}
\label{tab:random_walks}
\begin{tabular}{|p{1.4cm}||p{1.2cm}||p{1.8cm}||p{1.2cm}||p{1.2cm}|}
\hline
Transition Probability & Classical Random walk & Diffusive Random walk & Physical Random walk & Multi-Net Random walk \\
\hline
$w_{ii}^{\alpha\alpha}$ & $\frac{D_{ii}^{\alpha\alpha}}{s_i^\alpha}$ & $\frac{s_{max} + D_{ii}^{\alpha\alpha} - s_i^\alpha}{s_{max}}$ & 0 & 0 \\
\hline
$w_{ii}^{\alpha\beta}$ & $\frac{D_{ii}^{\alpha\beta}}{s_i^\alpha}$ & $\frac{D_{ii}^{\alpha\beta}}{s_{max}}$ & 0 & 0 \\
\hline
$w_{ij}^{\alpha\alpha}$ & $\frac{a_{ij}^{\alpha\alpha}}{s_i^\alpha}$ & $\frac{a_{ij}^{\alpha\alpha}}{s_{max}}$ & $\frac{a_{ij}^{\alpha\alpha}}{s_{i,\alpha}}\frac{D_{ii}^{\alpha\alpha}}{S_{i,\alpha}}$ & $\frac{1}{\mathscr{L} * deg\big(i_\alpha\big)}$ \\
\hline
$w_{ij}^{\alpha\beta}$ & 0 & 0 & $\frac{a_{ij}^{\alpha\beta}}{s_{i,\beta}}\frac{D_{ii}^{\alpha\beta}}{S_{i,\alpha}}$ & $\frac{1}{\mathscr{L} * deg\big(i_\beta\big)}$ \\

\hline

\end{tabular}
\end{table}
\end{center}

%\vspace{-0.4in}
\subsubsection{$\textsf{Multi-Net}$:Distributed representations of multiplex networks algorithm}
%\vspace{-0.2in}
\begin{algorithm}
\caption{Distributed representations of multiplex networks algorithm}
\begin{algorithmic}[1]
\Procedure{LearnFeatures}{Multiplex graph $MG=\big(V_i, E_i\big)_{i=1}^\mathscr{L}$, feature dimension \textit{d}, iterations per node \textit{n}, walk length \textit{l}, node sequence size \textit{k}}
% \State $\alpha$ = AssignTransitionProbability($MG$)
% \State $\beta$ = AssignSwitchProbability($MG$)
% \State $MG^\prime=(V_i,E_i,\alpha,\beta)$
\State Set node\_walks to \textit{Empty}

\For{i=1 to $\mathscr{L}$}
	\For{j=1 to n}
		\For{nodes $v_i \in MG_i$}
			\State node\_sequence = MULTIWALK($MG,v_i,l$)
			\State Add node\_sequence to node\_walks
		\EndFor
	\EndFor
\EndFor

\State \textit{f} = StochasticGradientDescent(d,k,node\_walks)
\State \textbf{return} f
\EndProcedure
\end{algorithmic}

\begin{algorithmic}[1]
\Procedure{multiwalk}{Multiplex network $MG=\big(V_i, E_i\big)_{i=1}^\mathscr{L}$, Source node \textit{u}, Walk length \textit{l}}
\State Create list \textit{walk} with \textit{u}
	\For{iter=1 to \textit{l}}
		\State U=get last node from \textit{walk}
		\State $G_{curr}$ = RandomUniformSampleLayers($MG$)
		\State $C_U$ = getNeighbors($U,G_curr$)
		\State $v$ = RandomUniformSampleNeighbors($C_U$)
		\State Add \textit{v} to \textit{walk}
	\EndFor
	\State \textbf{return} walk
\EndProcedure
\end{algorithmic}
\label{alg:multinet}
\end{algorithm}
\vspace{-0.1in}
Algorithm~\ref{alg:multinet} gives a pseudocode of our proposed distributed representation learning of a multiplex network \textsf{Multi-Net}. The random walk results are biased due to the start vertex \textit{u} of the random walker. We perform random walk of length \textit{l} from each node \textit{$v_i$} from each layer \textit{$MG_i$}(where $i=1,2, \ldots \mathscr{L}$) of the multiplex network \textit{MG}. We iterate over this process \textit{n} times to overcome the bias involved in randomness of choosing a node sequence. Thus, say if a node $v$ appears in two networks $MG_1,MG_2$, we collect context of the node($v$) $2*n$ times, $n$ times from node $v$ of $MG_1$ and $n$ times from node $v$ of $MG_2$, each time collecting a context or node sequence of length $l$ for the node $v$. During each random walk step, our algorithm chooses two random options: one for choosing next layer and one for choosing next node. We select next layer and node uniformly random. Thus each layer and node has equal transition probability to be chosen as next. The learned $d$ sized feature vector is optimized using a stochastic gradient descent (SGD).

%We assign transition probability($\alpha$), to visit the next neighbor, for each node based on their degree and switch probability($\beta$) for each to switch to another network layer based on number of layers. Given these transition probabilities, a random walk can be done in \textit{O(1)} using alias sampling. 
%\vspace{-0.2in}
\subsubsection{Use Cases}
Node embeddings from multiple networks using \textsf{Multi-Net} can be used inturn for variety of large scale network analytics tasks. Some of them are discussed below:

\subsubsection*{Node classification}
Given a network $G=\big( V,E \big)$ with a set of vertices $V$ and their corresponding labels $Y$, node classification is learning a mapping \textit{M}:$V \rightarrow Y$. The node embedding extracted from other layers of the multilayer network using \textsf{Multi-Net} or other ebedding techniques \cite{perozzi2014} \cite{grover2016} \cite{zitnik2017} can be given as input to a classifier like SVMs, Neural Networks, such that the classifier learns node features and their corresponding labels. With proper parameter tuning in the classifier and cross validations, we can optimize the parameters used for node embeddings models.

\subsubsection*{Network reconstruction}
%\vspace{-0.15in}
We define a network reconstruction problem as: given a multiplex network $MG=\big(V_i, E_i\big)_{i=1}^\mathscr{L}$ with $\mathscr{L}$ layers and vector representations of $\mathscr{V}$ nodes of $\mathscr{L}-1$ layers, network reconstruction task is a boolean classifier task that learns a mapping \textit{N}:$\big(\mathscr{E}_{\mathscr{L}}\big) \rightarrow \{0|1\}$ of the network using the node embeddings and predict edges and non-edges of the entire $\mathscr{L}^{th}$ layer. The training of a boolean classifier is made on edges($\mathscr{E}_{\mathscr{L}}$) and non-edges($\mathscr{E}^\prime_{\mathscr{L}}$) of the $MG_{\mathscr{L}^{th}}$ layer in the network.  We aim on network reconstruction task because in most of the social network streams, getting the underlying social network is a tedious process(ex. Twitter API has very limited accessibility for their \textit{follower network}).
%\vspace{-0.2in}
\section{Experiments and Results}
%\vspace{-0.2in}
\subsection{Datasets}
We use 4 datasets to evaluate our approach of multiplex network embedding. 
%1. \textbf{Caenorhabditis Elegans wiring network}~\cite{chen2006,de2015} is a 3-layer network where each layer corresponds to synaptic junction in the Caenorhabditis Elegans connectome, 2. \textbf{Homo Genetic network}~\cite{stark2006,de2015} is a 4-layer homo sapiens gene interaction networks, 3. \textbf{Virtual World}~\cite{boccaletti2014} is a 4-layer online gaming network that represents various interactions of users within the virtual world, 4. \textbf{Twitter network} is a 4-layer twitter network~\cite{de2013} collected during Higgs Boson event, where each layer represent different interaction types of Twitter users. Properties of all datasets are specified in Table~\ref{tab:data_prop}. 

\subsubsection*{Caenorhabditis Elegans wiring Network}
This neuronal multiplex network comprise of 3 layers: \textit{electric}, \textit{chemical poladic} and \textit{chemical monadic}, where each layer corresponds to synaptic junction in the Caenorhabditis Elegans nematode~\cite{chen2006,de2015}

\subsubsection*{Homo Genetic Network}
We collected a sample of gene interaction data of homo sapiens from BioGRID~\cite{stark2006} and CoMuNe lab~\cite{de2015}. The original data contains multiplex network with 7 layers. Since most of the nodes in that network are present only in one network, we took a sample of that data with 4 layers of genetic interaction: \textit{direct interaction}, \textit{physical association}, \textit{association} and \textit{colocalization}.

\subsubsection*{Gaming network}
We collected a 4-layer multiplex network that encompasses various social interactions within a virtual environment~\cite{boccaletti2014} made available in Harvard Dataverse. The interactions include: \textit{friendship}, \textit{messaging}, \textit{transactions} and \textit{visits}. Similar to Homo Genetic network, we sampled a part of this network to remove isolated nodes and nodes that exist in only one layer.

\subsubsection*{Twitter network}
For our initial analysis, we utilized open access data of Twitter feed networks \cite{de2013}. This data comprise of 4 layered multilayer network from Twitter, that are collected during the Higgs Boson event. Each layer represents Twitter's reply, mention, retweet and social relationship network. Due to copyright issues, all personal information like user names, id, country, religion, etc are hidden from the data. The nodes are represented by node ID's of $1,2,...,n$. The entire data houses approximately 456,000 unique nodes across all layers. Table~\ref{tab:data_prop} gives some basic properties of all layers in the given multilayer network. 

\begin{table}
\centering
\caption{Properties of datasets}
\label{tab:data_prop}
\begin{tabular}{|p{1.8cm}||p{2.1cm}||p{2.1cm}||p{2.1cm}||p{2.1cm}|}
%\begin{tabular}{|c||c||c||c||c|}
\hline
 & C. Elegans & H. Genetic & Virtual World & Higgs Boson \\
\hline
Layers & Electric junction, Poladic and Monadic junctions & Direct \& Physical Associations, Association, Colocalization & Friendship, Messaging, Transactions, Visits & Reply, Retweet, Mention, Friendship \\
\hline
Link prediction & Electric junction & Colocalization & Friendship & Friendship \\
\hline
\# of nodes & 279 & 17,927 & 39,832 & 456,626 \\
\hline
\# of edges & 5,863 & 169,238 & 2,683,743 & 15,361,596 \\
\hline
Avg. clustering coefficient & 0.22 & 0.14 & 0.09 & 0.07 \\
%\hline
%Avg. \# of connected components & 1 & 90 & 1 & 1 \\
\hline
\hline

\end{tabular}
\end{table}

\subsection{Experimental Setup}
%\vspace{-0.2in}
We evaluate the proposed feature learning methods with the network reconstruction task. We compare and contrast results obtained from the proposed method with the following deep learning based feature learning methods, with parameters same as in their proposed work:

\begin{itemize}
	\item \textbf{DeepWalk} \cite{perozzi2014}: This method extract \textit{d-dimensional} features based on uniform random walks
	\item \textbf{Node2Vec} \cite{grover2016}: This method learns \textit{d-dimensional} features using biased random walk strategy which helps in flexible exploration of neighborhood. We set the parameters for exploring the neighborhood nodes: \textit{p} and \textit{q} as 1
	\item \textbf{OhmNet} \cite{zitnik2017}: This approach uses \textsf{Node2Vec} method for feature learning, but for a multilayer network setup. This approach defines a hierarchy for representing nodes from multiple layers. We use same parameters as \textsf{Node2Vec} and we create a 2-level hierarchy for all multiplex networks.
	\item \textbf{MNE} \cite{zhang2018}: This latest approach is similar to \textsf{Node2Vec}. This approach considers a base network of a multiplex network to learn the embedding of all nodes and individual representation for each layer. Since this approach uses \textsf{Node2Vec} as a base method, we use same parameters as \textit{Node2Vec} algorithm.
\end{itemize}

Since \textsf{DeepWalk} and \textsf{Node2Vec} methods are suitable for single layer networks in nature, in our experiments, we collapse $\mathscr{L}$ layers in the multiplex network into a single network. A disadvantage of this approach is that it loses huge amount of structural dependencies of nodes belonging to each layer and may give unstable prediction scores. 

%Since \textsf{DeepWalk} and \textsf{Node2Vec} methods are suitable for single layer networks in nature, we experiment on two different feature representation for multilayer networks for both \textsf{DeepWalk} and \textsf{Node2Vec}:

%\begin{itemize}
%	\item Independent layers: In this method, we learn feature representations of one layer at a time using \textsf{DeepWalk} or \textsf{Node2Vec} and retrain feature learning model for nodes in other layer of networks
%	\item Collapsed layers: In this approach, we merge all layers of networks into a single network before feature learning. We then use \textsf{DeepWalk} or \textsf{Node2Vec} to learn node features from the merged network.
%\end{itemize}

For all multiplex networks with $\mathscr{L}$, we learn representations of $\mathscr{L}-1$ layers and perform network reconstruction task on $\mathscr{L}^{th}$ layer using the \textit{L2-regularized Logistic Regression}. Refer Table~\ref{tab:data_prop} for getting details on which layer we set for link prediction task. Further, we set same parameter values(\textit{walk length (\textit{l}=10)}, \textit{number of walks per node (\textit{r}=5)}, \textit{feature dimension (\textit{d}=150)}, \textit{context window (\textit{w})=10}) for all experiments.

\subsection{Experiment results}

%\vspace{-0.2in}
%From multiple layers of networks, we learn properties of nodes in a low-dimensional vector space. We use these representations to build a model for performing link prediction task. We use logistic classifier with L2 regularization and we use random 50\% of edges and same ratio of non-edges from the friends/follower network. 
%As mentioned in Section~\ref{subsec:friend_net}, we use multiple networks of the Twitter data to learn representations of users and use such representations to construct the base friends network of Twitter. 
We compare results of state-of-the-art methods~\cite{perozzi2014,grover2016,zitnik2017,zhang2018} to evaluate \textsf{Multi-Net}. Due to space constraints we limit our analysis only based on \textsf{Collapsed DeepWalk}, \textsf{Collapsed Node2Vec}, \textsf{OhmNet}, \textsf{MNE} and \textsf{Multi-Net} with multiple random walk strategies. Since the single layer based \textsf{Node2Vec} and \textsf{DeepWalk} provided marginal performance, we do not include their results for analysis. 

%\vspace{-0.2in}
\begin{figure*}[!ht]
	\centering
		\subfloat[Figure a: Results of small networks\label{fig:small}]{\includegraphics[scale=0.35]{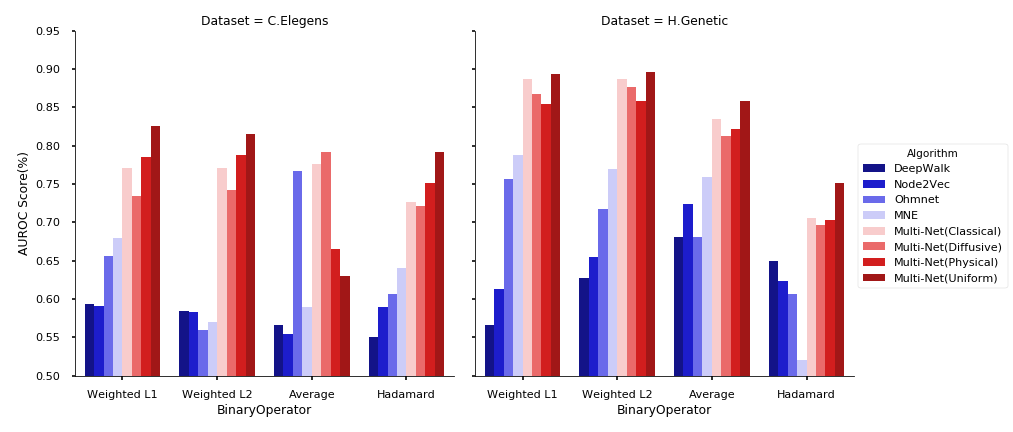}}
		
		\qquad
		\subfloat[Figure b: Results of larger networks\label{fig:big}]{\includegraphics[scale=0.35]{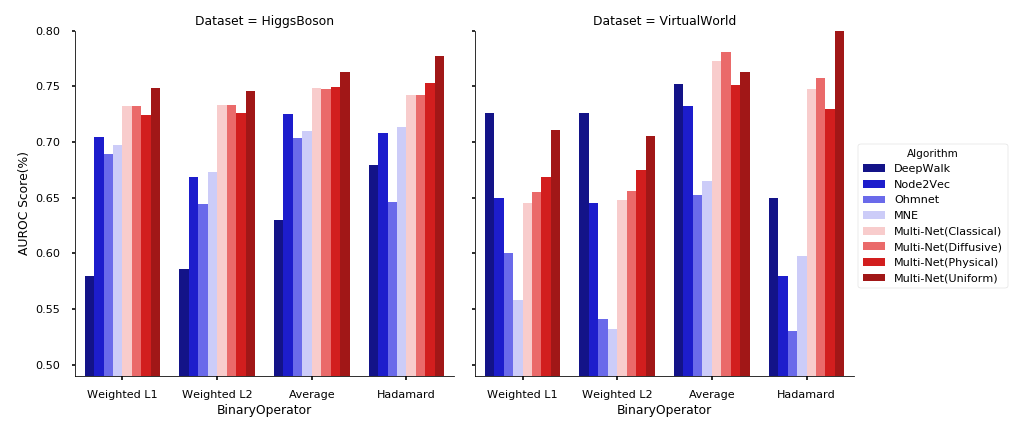}}

		\caption{AUROC scores in ($\%$) for all models used in experiments with different binary operators to join a node's embeddings. Red bars represent results of random walks given in this paper and Blue bars represent results of models from the literature. \textbf{Multi-Net(Uniform)} is the proposed random walk method.}
		\label{fig:results}
\end{figure*}

\begin{table}
\caption{Binary operators for getting edge features from features of nodes \textit{u} and \textit{v}($f\big(u\big)$ and $f\big(v\big)$ respectively)}
\centering
\begin{tabular}{|p{4cm}||p{4cm}|}
\hline
\textbf{Binary Operator} & \textbf{Formula}\\
\hline
Hadamard & $[f\big(u\big).f\big(v\big)] = f\big(u\big) * f\big(v\big)$\\
\hline
Average & $[f\big(u\big)+f\big(v\big)] = \frac{f\big(u\big) + f\big(v\big)}{2}$\\
\hline
Weighted L1 & $||f\big(u\big).f\big(v\big)||_1 = |f\big(u\big) - f\big(v\big)|$\\
\hline
Weighted L2 & $||f\big(u\big).f\big(v\big)||_2 = |f\big(u\big) - f\big(v\big)|^2$\\
\hline
\end{tabular}
\label{tab:operators}
\end{table}
%\vspace{-0.2in}
We conducted all experiments using 4 binary operators: \textit{Hadamard, Average, Weighted L1, Weighted L2} for getting vector representation of an edge. Formula of these binary operators are given in Table~\ref{tab:operators}. 

In Figure~\ref{fig:results}, we give plots of Area Under ROC (AUROC) curve score for different models over different binary operators and datasets. We give plots of small datasets(\textit{C.Elegans} \& \textit{H.Genetic}) in Figure~\ref{fig:small} and plots of bigger datsaets(\textit{Higgs Boson} \& \textit{Virtual World}) in Figure~\ref{fig:big}. We can note that models with multilayer random walks performing better than the state-of-the-art models, because of multilayer random walks effectively capturing context of nodes across all layers in the given multiplex network. In majority of the cases, we can note that \textsf{Multi-Net} on top of the proposed simple random walk strategy gives atleast 11\% better results than existing state-of-the-art models and atleast 5\% improvement over other random walk methods. 

In relation with binary operators of combining embedding of nodes, we note Multi-Net performs better with \textit{Weighted L1} and \textit{Weighted L2} operators in smaller networks(Figure~\ref{fig:small}) with around $10\%$ performance improvement over state-of-the-art on average. On the other hand, \textit{Hadamard} operator suits \textsf{Multi-Net} over the proposed random walk for larger networks(Figure~\ref{fig:big}) with around $12\%$ performance gain over the state-of-the-art method multiplex network embedding methods. 

%We can note from these plots that Multi-net method on top of multiple multilayer random walks performed better than state-of-the-art network representation methods. In Figure~\ref{fig:acc_roc}, we give all model evaluations based on model accuracy, Area Under ROC (AUROC) curve score. With these evaluation, we can note that the proposed \textit{Multi-Net model} outperforms \textsf{DeepWalk} and \textsf{Node2Vec} based models by around \textit{7\%} increase. Most notably, it outperforms \textsf{OhmNet} model by \textit{12\%} on average.

%In Figure~\ref{fig:prc_f1}, we evaluate all models using Area Under Precision Recall Curve (AUPRC) score and average \textit{F1} measure. From these plots, we can note that Multi-Net models with proposed random walk strategies performed better with respect to AUPRC and F1 scores. Although all model performances are equal in terms of AUPRC, we can find significant improvement with F1 measure with Weighted L2 operator and Hadamard operator. It is notable that \textsf{Multi-Net} achieves around \textit{5\%} gain over \textsf{DeepWalk} and \textsf{Node2Vec} based models with F1 measure. Again, the \textsf{Multi-Net} model outperforms \textsf{OhmNet} by around \textit{13\%} gain.

%From Figure~\ref{fig:results}, we can note that \textsf{Multi-Net} performs better with \textit{Hadamard} and \textit{Weighted L2} operators, while \textsf{Node2Vec} and \textsf{OhmNet} performs better with \textit{Average} binary operator. In all binary operators, the \textsf{Multi-Net model} performs better than other models with approximately 7\% gain.

From our experiments, we also note that, in some cases, simple \textsf{Node2Vec} and \textsf{DeepWalk}, outperforms the \textsf{OhmNet} and \textsf{MNE} models, which are intended to perform better for multiplex networks. This is may be due to the fact that the models requires careful engineering on parameters(\textit{hierarchies} in \textsf{OhmNet} and \textit{embedding dimensions} in \textsf{MNE}). From our extensive analysis, we can show that the proposed multilayer random walk method helps to get better embedding of nodes, which improves prediction scores. The given random walk strategies can be applied to existing models like MNE to improve their prediction scores. 

%\vspace{-0.2in}
\subsection{Evaluation on algorithm scalability}
\begin{figure}
	\centering
	\includegraphics[scale=0.5]{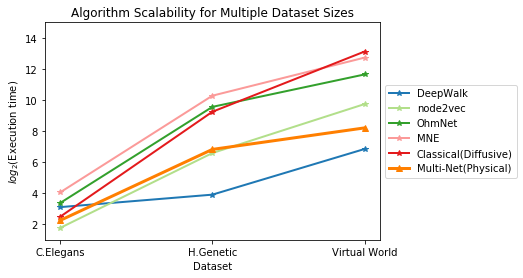}
	\caption{Execution times of multiple algorithms on small-medium sized datasets. Methods in parenthesis denote that it has execution time almost similar to methods outside parenthesis}
	\label{fig:scalability}
\end{figure}
%\vspace{-0.2in}
In Figure~\ref{fig:scalability}, we give execution times of all models on different datasets. We can see from this figure that \textsf{Multi-Net} and \textsf{Physical} random walk procedures are more scalable compared to existing state-of-the-art multiplex network embedding methods. We also note that Classical and Diffusive random walk methods performs better only if the network size is very small and the performance degrades when number of edges increase in the network. Thus the scalability of these two random walk procedures are lower compared to state-of-the-art methods. This is due to the overhead of extracting shortest paths between nodes across multiplex networks. Eventhough, Multi-Net and Physical random walks performs poor compared to \textsf{DeepWalk}, their performance improves for larger networks like Twitter. From Figures~\ref{fig:results} and ~\ref{fig:scalability}, we can note that the proposed Multi-Net method using the proposed random walk performs better in terms of accuracy and scalability.

%\vspace{-0.3in}
\section{Discussion and Future Work}
%\vspace{-0.2in}
The crux of developing network embeddings given a graph is to design a random-walk procedure. Moreover, in a multiplex network we need an effective way to aggregate the low-dimensional node embeddings across the different layers. In this work, we presented \textsf{Multi-Net}: a basic framework on learning features in an unsupervised way from a multilayer network as an optimization problem. We have given a simple search strategy on the multiplex network to efficiently get a node's context or neighborhood in a multiplex network setup. Our method benefits from the \emph{occam's razor} principle for designing random walk procedures by assigning transition probability to a node and a simple switching probability across layers. Particularly, the effectiveness of \textsf{Multi-Net} in reconstructing a multiplex network layer from representations of nodes from other layers in the network.

This research has a lot of scope for future explorations. For example, at present there are approaches available to learn feature representations based on static networks. But in reality, networks evolve over time and the topology of a node in multiple layers change over a span of time \cite{paranjape2017}. Temporal networks can effectively model data from multiple fields like financial transactions and social media cascades. Another addition to the future work can be considering higher order motifs and graphlets \cite{benson2016}. All existing network embedding techniques focus on pair wise relationship among nodes, compromising the actual structural values of a network. However, motifs and graphlets by default captures such properties.
%\vspace{-0.2in}

\bibliographystyle{spmpsci}
{\footnotesize
\bibliography{bibliography}}

\end{document}